\documentstyle[aps,prb,preprint]{revtex}

\begin{document}
\title{Ferromagnetic tunneling junctions at low voltages:\\
elastic {\em versus }inelastic scattering at $T=0^{\circ }K$.}
\author{C. A. Dartora and G. G. Cabrera}
\address{Instituto de F\'{\i }sica `Gleb Wataghin',\\
Universidade Estadual de Campinas (UNICAMP),\\
C. P. 6165, Campinas 13083-970 SP, Brazil}
\date{\today}
\maketitle

\begin{abstract}
In this paper we analyze different contributions to the magnetoresistance of
magnetic tunneling junctions at low voltages. A substantial fraction of the
resistance drop with voltage can be ascribed to variations of the density of
states and the barrier transmission with the bias. However, we found that
the {\em anomaly} observed at zero bias and the magnetoresistance behavior
at very small voltages, point to the contribution of inelastic
magnon-assisted tunneling. The latter is described by a transfer parameter $%
T^{J}$, which is one or two orders of magnitude smaller than $T^{d}$, the
direct transmission for elastic currents. Our theory is in excellent
agreement with experimental data, yielding estimated values of $T^{J}$ which
are of the order of $T^{d}/T^{J}\sim 40$.
\end{abstract}



\narrowtext
\tightenlines

\section{Introduction}

Recently, the interest on the phenomena of Giant Magnetoresistance (GMR) in
magnetic tunnel junctions (MTJ) has grown significantly due to potential
applications in magnetoresistive reading heads, magnetic field sensors,
nonvolatile magnetic random access memories, and many others\cite
{[1],[2],[3],[4],[5]}. The effect is based on the spin dependent scattering
mechanisms proposed in the early papers by Cabrera and Falicov\cite{[18]},
which lead in MTJ's, to a strong dependence of the conductance on the
magnetic polarization\cite{[6]}. Typically, the GMR effect found in MTJ's is
of the order of $25-30\%$\cite{[7],moodera}, and points to a large ratio of
the densities of states for majority ($M$) and minority ($m$) electrons at
the Fermi level ($E_{F}$) 
\[
\frac{N_{M}\left( E_{F}\right) }{N_{m}\left( E_{F}\right) }\approx 2.0-2.5\
. 
\]
As usual in MR experiments, one compares the resistances for the cases where
the magnetizations at the electrodes are anti-parallel (AP) and parallel
(P). In several experiments reported in the literature (see for example Ref. %
\onlinecite{[1],[2],[7],moodera}), the junction resistance drops
significantly with the applied voltages, with a sharp peak at zero bias (%
{\em zero-bias anomaly}). This bias dependence shows a rapid initial
decrease up to voltages of the order of $V\sim 100\ mV$, then slows down but
continues decreasing with voltages, up to $60\%$ of the peak value at $500\
mV$ in some cases \cite{moodera}. Many attempts to explain the above
behavior have been done over the last years \cite{[1],[2],[7],[8]}, but a
complete theory which includes all the observed features is still lacking.

In Ref. \onlinecite{[7]}, scattering from magnons at the electrode-insulator
interface has been proposed as the mechanism for randomizing the tunneling
process and opening the spin-flip channels that reduce the MR. While this
process may explain the MR behavior in the vicinity of zero-bias (voltages
smaller than $40-100\ mV$), estimations of magnon scattering cross sections
show that the effect is too small to account for the sharp drop in
resistance observed in the whole range of $500\ mV$. In fact,
inelastic-electron tunneling spectroscopy (IETS) measurements at low
temperature showed peaks which can unambiguously be associated with
one-magnon spectra at very small voltages (from $12$ to $20mV$, with tails
up to $40mV$, and maximum magnon energy not larger than $100$ $meV$) \cite
{[1]}. To go beyond this limit will imply multimagnon processes, which are
negligible at low temperature. This way, the electron-magnon coupling
constant coming from Ref. \onlinecite{[7]} is by sure considerable
overestimated.

The above explanation\cite{[7]} has been challenged in Ref. \onlinecite{[8]}%
, where it is shown that the experimental data can be understood in terms of
elastic tunneling currents which conserve spin, by considering effects not
taken into account in Ref. \onlinecite{[7]}. Those include the lowering of
the effective barrier height with the applied voltage, as in the classical
Simmons' theory\cite{[10]}, and most important, variations of the densities
of states with the bias at both magnetic electrodes. The latter is a
relevant question, since experiments probe depths of the order of $0.5$ $eV$
from the Fermi surface. The simple calculation developed in Ref. %
\onlinecite{[8]} models the band structure with free electron-like densities
of states, since the tunneling current is dominated by the {\em s-}electron
contribution. This approach yields a {\em zero bias} anomaly which depends
on the band structure, and a variation of the MR which has the right order
of magnitude for the whole range of $500$ $meV$. The above discussion and
other experimental results primarily exhibit that the density of states
dependence on the applied voltage plays an important role \cite{other,[11]}.
However, fine details of experiments at very small voltages are difficult to
fit. One may adopt here a pragmatic procedure, with a more intricate band
structure and more free parameters to improve the fitting\cite{[2]}.

In this paper we take a different stand, motivated by results from IETS
experiments\cite{[1]}, which show that inelastic scattering do participate
in the phenomenon at very small voltages. Also, MR experiments\cite
{[7],moodera} show clearly a different behavior with applied voltage in the
same small bias region (up to $100$ $mV$). A complete theory then should
include:

\begin{enumerate}
\item[{\em i)}]  magnon assisted tunneling effects, with maximum magnon
energies of the order of $\sim 100$ $meV$. At low temperature, electrons
from the electrodes, accelerated by the applied voltage, excite magnons at
the interface. At low temperature, only magnon-emission processes should be
considered;

\item[{\em ii)}]  variation with voltages of the densities of states for the
different spin bands in the ferromagnets. Here, we will follow closely the
approach of Ref. \onlinecite{[8]}, with a simple picture of the band
structure. This is motivated by the discussions given in Ref. %
\onlinecite{nico} and \onlinecite{[8]} over the polarization of the
tunneling current. We assume here that the latter is mainly of {\em s}%
-character;

\item[{\em iii)}]  lowering of the effective barrier height with the applied
voltage. This effects, as shown in Ref. \onlinecite{[10]}, yields to a
parabolic dependence of the resistance with the bias. It does not contribute
to the {\em zero-bias }\ anomaly, but it is always present and should
dominate the behavior at large voltages.
\end{enumerate}

The above program will be developed in the present contribution. The content
of this paper can be described as follows: In the next Section, we formulate
the theoretical basis for analyzing tunneling currents, discussing the
transfer Hamiltonian which includes all the above mentioned ingredients. In
Section 3, we solely analyze density of states effects, considering elastic
tunneling processes. Some analytical expressions are shown. In Section 4, we
include contributions from inelastic magnon-assisted processes to the
tunneling current, and finally, in the last Section, a few conclusions and
remarks are added.

\section{Theoretical Preliminaries}

To give a description of the MR and the resistance in the MTJ, we will use
the transfer Hamiltonian method \cite{[12]}. The junction is composed by two
ferromagnetic electrodes separated by a thin oxide film which represents a
potential barrier due to the fact that the Fermi levels of the ferromagnetic
layers are situated in the gap region of the oxide film. We have considered
the {\em s}-band electrons as free particles (plane-waves), being
responsible for the dominant contribution to the tunneling process. The {\em %
d} electrons, which are more localized, enter in the process only via the
exchange interaction with {\em s} electrons on each ferromagnetic electrode.
In the context of second quantization and neglecting the magnetization
energy (Zeeman term), the unperturbed Hamiltonian is given by 
\begin{equation}
H_{0}=\sum_{{\bf k}\sigma ,\alpha =(L,R)}E_{{\bf k}\sigma }c_{{\bf k}\sigma
}^{\alpha \dagger }c_{{\bf k}\sigma }^{\alpha }  \label{H0}
\end{equation}
with $L(R)$ referring to the left (right) ferromagnetic electrode, $c_{{\bf k%
}\sigma }^{\alpha \dagger }$ ($c_{{\bf k}\sigma }^{\alpha }$) are the
creation (annihilation) fermionic operators for wave vector ${\bf k}$ and
spin $\sigma $, $E_{{\bf k}\sigma }=\frac{\hbar ^{2}k^{2}}{2m}-\sigma \Delta
_{\alpha }$ is the Hartree-Fock energy and $\Delta _{\alpha }$ is the shift
in energy due to exchange interaction in each side of the barrier.

In writing the interaction part of the total Hamiltonian, which makes
possible the transfer of electrons from one side of the insulating barrier
to the other, we follow Ref.\onlinecite{[7]}. Apart from the direct transfer
which comes from elastic processes, we include transfer with magnetic
excitations that originates from the {\em s-d }exchange between conduction
electrons and localized spins at the interfaces. The excitations are
described by a linearized Holstein-Primakoff transformation \cite{[15]}, in
the spirit of a one-magnon theory. We use the following Hamiltonian:

\begin{eqnarray}
H_{int} &=&\sum_{{\bf kk}^{\prime }\sigma }\ t_{{\bf kk}^{\prime
}}^{d}\left( c_{{\bf k}\sigma }^{L\dagger }c_{{\bf k}^{\prime }\sigma
}^{R}+c_{{\bf k}^{\prime }\sigma }^{R\dagger }c_{{\bf k}\sigma }^{L}\right) +
\label{hint} \\
&&+\frac{1}{\sqrt{N_{s}}}\sum_{{\bf kk}^{\prime }{\bf q}}\ t_{{\bf kk}%
^{\prime }{\bf q}}^{J}\left( c_{{\bf k}\downarrow }^{L\dagger }c_{{\bf k}%
^{\prime }\uparrow }^{R}+c_{{\bf k}^{\prime }\downarrow }^{R\dagger }c_{{\bf %
k}\uparrow }^{L}\right) \left( \sqrt{2S_{L}}b_{{\bf q}}^{L}+\sqrt{2S_{R}}b_{%
{\bf q}}^{R}\right) +  \nonumber \\
&&+\frac{1}{\sqrt{N_{s}}}\sum_{{\bf kk}^{\prime }{\bf q}}\ t_{{\bf kk}%
^{\prime }{\bf q}}^{J}\left( c_{{\bf k}\uparrow }^{L\dagger }c_{{\bf k}%
^{\prime }\downarrow }^{R}+c_{{\bf k}^{\prime }\uparrow }^{R\dagger }c_{{\bf %
k}\downarrow }^{L}\right) \left( \sqrt{2S_{L}}b_{{\bf q}}^{L\dagger }+\sqrt{%
2S_{R}}b_{{\bf q}}^{R\dagger }\right) +  \nonumber \\
&&+\frac{1}{N_{s}}\sum_{{\bf kk}^{\prime }{\bf q}}\ t_{{\bf kk}^{\prime }%
{\bf q}}^{J}\left( c_{{\bf k}\uparrow }^{L\dagger }c_{{\bf k}^{\prime
}\uparrow }^{R}-c_{{\bf k}\downarrow }^{L\dagger }c_{{\bf k}^{\prime
}\downarrow }^{R}+h.c.\right) \left( S_{L}+S_{R}-\left( b_{{\bf q}%
}^{R\dagger }b_{{\bf q}}^{R}+b_{{\bf q}}^{L\dagger }b_{{\bf q}}^{L}\right)
\right) \ ,  \nonumber
\end{eqnarray}
where $t_{{\bf kk}^{\prime }}^{d}$ is the direct transmission coefficient, $%
t_{{\bf kk}^{\prime }{\bf q}}^{J}$ is the inelastic transmission coefficient
(depends on the exchange integral), $S^{L}$($S^{R}$) is the spin value at
the left (right) side, $N_{s}$ is the total number of spins at the
interface, and $b_{{\bf q}}^{\alpha \dagger }$ ($b_{{\bf q}}^{\alpha }$) are
the creation (annihilation) operators for magnons with wave vector ${\bf q}$
at each interface between the barrier and the electrodes. The wave vector $%
{\bf q}$ is quasi-two dimensional (the magnon wave function is localized at
the interfaces, but with finite localization length).

In general, the total current obtained with (\ref{hint}) has contributions
from elastic processes, resulting in a direct tunneling which conserves
spin, and from the inelastic ones, which involve emission and absorption of
magnons with electronic spin flip. In the following we describe the direct
term.

\section{Direct tunneling current: analytical expressions}

Considering only the direct part of the tunneling process, which means
elastic processes, without involving magnon excitations, the current is
easily obtained by \cite{[2],[3],[7],[8]}

\begin{equation}
I_{(C)}=\frac{2\pi e}{\hbar }\int dE~T^{d}(E,V,d,\Phi
_{0})~W_{(C)}(E,V)[f(E-eV)-f(E)]  \label{Id1}
\end{equation}
where 
\begin{equation}
W_{(C)}=\sum_{\sigma }N_{\sigma }^{R}(E)N_{\sigma }^{L}(E-eV)  \label{Id2}
\end{equation}
and $C$ denotes the configuration scheme, $C=P$ for parallel and $C=AP$ for
anti-parallel, $f(E)$ is the Fermi-Dirac distribution, $N_{\sigma }^{R}$ and 
$N_{\sigma }^{L}$ the density of states at the right and left electrodes,
respectively. $T^{d}(E,V,d,\Phi _{0})=|t_{kk^{\prime }}^{d}|^{2}$ is the
tunneling coefficient, being a function of the energy $E$, the applied
voltage $V$, the thickness of the barrier $d$ and the barrier height $\Phi
_{0}$. In fact, $T^{d}$ is a function of the overlap integral between the
left and right wave functions inside the barrier region.

The resistance is readily obtained by $R=G^{-1}$, where $G=dI/dV$ is the
differential conductance. In the low bias regime, we are interested in
voltages smaller than the Fermi energy and only the states near the Fermi
level will contribute to the transport, so we can expand the density of
states in a Taylor series as follows:

\begin{equation}
N_{\sigma }^{\alpha }(E)=\sum_{n=0}^{\infty }\frac{1}{n!}\left. \frac{%
d^{n}N_{\sigma }^{\alpha }(E)}{dE^{n}}\right| _{E_{F}}(E-E_{F})^{n}~~.
\label{Taylor}
\end{equation}

Now, let us calculate $W_{C}$ for the $P$ and $AP$ configurations, using (%
\ref{Taylor}). In the $P$ configuration the majority and minority bands in
each electrode corresponds to the same spin orientation, and in $AP$
configuration the majority band of one electrode is the minority on the
other: 
\begin{eqnarray}
W_{(P)} &=&\sum_{i}\sum_{j}\frac{1}{i!j!}\left[ \frac{d^{i}N_{M}^{R}\left(
E\right) }{dE^{i}}\frac{d^{j}N_{M}^{L}\left( E-eV\right) }{d\left(
E-eV\right) ^{j}}\right. +  \label{conf1} \\
&&+\left. \left. \frac{d^{i}N_{m}^{R}\left( E\right) }{dE^{i}}\frac{%
d^{j}N_{m}^{L}\left( E-eV\right) }{d\left( E-eV\right) ^{j}}\right] \right|
_{E_{F}}\left( E-E_{F}\right) ^{i}\left( E-eV-E_{F}\right) ^{j}\ ,  \nonumber
\end{eqnarray}
and

\begin{eqnarray}
W_{(AP)} &=&\sum_{i}\sum_{j}\frac{1}{i!j!}\left[ \frac{d^{i}N_{M}^{R}\left(
E\right) }{dE^{i}}\frac{d^{j}N_{m}^{L}\left( E-eV\right) }{d\left(
E-eV\right) ^{j}}\right. +  \label{conf2} \\
&&+\left. \left. \frac{d^{i}N_{m}^{R}\left( E\right) }{dE^{i}}\frac{%
d^{j}N_{M}^{L}\left( E-eV\right) }{d\left( E-eV\right) ^{j}}\right] \right|
_{E_{F}}\left( E-E_{F}\right) ^{i}\left( E-eV-E_{F}\right) ^{j}\ ,  \nonumber
\end{eqnarray}

Taking into account identical electrodes and the low bias regime, we can
expand these expressions to first order with good accuracy. The {\em s}-band
can be represented by a parabolic dispersion relation and density of states $%
N_{\sigma }\propto \sqrt{E-\Delta _{\sigma }}$, where $\Delta _{\sigma
}(\sigma =\uparrow ,\downarrow )$ gives the bottom of the spin band, with $%
\left| \Delta _{\uparrow }-\Delta _{\downarrow }\right| =2\Delta ,$ as in
Ref. \onlinecite{[8]}. However, we consider here cases more general than the
parabolic dispersion, with the band structure described through the
following set of parameters: 
\begin{equation}
\begin{array}{l}
r\equiv \left( 
{\displaystyle {N_{M} \over N_{m}}}%
\right) _{F}\ , \\ 
\lambda \equiv \left( 
{\displaystyle {dN_{M}/dE \over dN_{m}/dE}}%
\right) _{F}\ , \\ 
\beta \equiv \left( 
{\displaystyle {1 \over N_{m}}}%
{\displaystyle {dN_{m} \over dE}}%
\right) _{F}\ ,
\end{array}
\label{parameters}
\end{equation}
with all quantities evaluated at the Fermi level, and $m$ and $M$ stand for
minority and majority spin bands, respectively. We get the analytic
expressions:

\begin{equation}
W_{(P)}=(N_{m}^{F})^{2}\left\{ \left( 1+r^{2}\right) +\beta \left(
1+r\lambda \right) (2\varepsilon -V)+\beta ^{2}\left( 1+\lambda ^{2}\right)
\varepsilon (\varepsilon -V)\right\} \ ,  \label{parallel}
\end{equation}
and

\begin{equation}
W_{(AP)}=(N_{m}^{F})^{2}\left\{ 2r+\beta \left( r+\lambda \right)
(2\varepsilon -V)+\beta ^{2}\lambda \varepsilon (\varepsilon -V)\right\} \ ,
\label{antiparallel}
\end{equation}
where $\varepsilon \equiv E-E_{F}$ and $\varepsilon $ and $V$ must be given
in $eV$.

There are several possibilities for including the tunneling transmission
coefficient $T^{d}$ in the theory. One is the approach followed by Simmons 
\cite{[10]}, where the barrier is parametrized by an effective height $\Phi
_{0}$ and an effective thickness $d$: 
\begin{equation}
T^{d}(E,V,\Phi _{0},d)=\exp [-\frac{2}{\hbar }d\sqrt{2m(\Phi
_{0}-\varepsilon _{z})}]\ =\exp \left[ -1.024\ d\sqrt{\Phi _{0}}\right] \exp
\left[ \frac{1}{2}\frac{\eta \epsilon }{\sqrt{\Phi _{0}}}\right]
\label{transm}
\end{equation}
where all energies are measured from the Fermi level and given in $eV$, the
barrier width given in $Angstrom=0.1\ nm$, and $\eta $ is some constant
relating the energy $\varepsilon $ with its component $\varepsilon _{z}$
perpendicular to the barrier. This latter parameter appears due to the fact
that we are using a one dimensional formula to explain the behavior in the $%
3D$ case.

Since the Fermi-Dirac functions are step-like at $0^{\circ }K$, we can
easily obtain the conductance for both configurations. 
\[
G_{\left( C\right) }=\frac{2\pi e^{2}}{\hbar }\frac{d}{dV}\left\{
\int_{0}^{V}d\epsilon \ T^{d}(\varepsilon ,V,\Phi _{0},d)\ W_{(C)}\left(
\varepsilon ,V\right) \right\} 
\]
$\ .$

With some simplifications in the integration process (taking into account
the behavior of the integrand in the range of integration, and making use of
some geometric arguments), one obtains 
\begin{eqnarray}
G_{\left( C\right) } &=&\frac{2\pi e^{2}}{\hbar }\left\{ A_{(C)}T^{d}(V,\Phi
_{0},d)\ +\right.  \label{gc} \\
&&\left. +\frac{1}{6}\frac{d}{dV}\left[ B_{(C)}V^{2}T^{d}(V,\Phi
_{0},d)-C_{(C)}V^{3}T^{d}(3V/5,\Phi _{0},d)\right] \right\} \ ,  \nonumber
\end{eqnarray}
being $A_{(C)},B_{(C)}$ and $C_{(C)}$ constants related to the configuration
scheme and the density of states. Following, the analytical expressions for
the conductance in both parallel and anti-parallel configurations are
presented, using (\ref{gc}) and considering the expansions (\ref{parallel})
and (\ref{antiparallel}):

\begin{eqnarray}
G_{(P)} &=&\frac{2\pi e^{2}}{\hbar }\exp [-1.024d\sqrt{\Phi _{0}}%
][N_{m}^{F}]^{2}\left\{ (1+r^{2})\exp [\frac{\eta Vd}{2\sqrt{\Phi _{0}}}]+%
\frac{\beta (1+r\lambda )}{3}\times \right.  \nonumber \\
&&\times \left[ \frac{\eta dV^{2}}{4\sqrt{\Phi _{0}}}\exp [\frac{\eta Vd}{2%
\sqrt{\Phi _{0}}}]+V(\exp [\frac{\eta Vd}{2\sqrt{\Phi _{0}}}]-1)\right] -
\label{gp} \\
&&\qquad \qquad \left. -\frac{\beta ^{2}(1+\lambda ^{2})}{2}\exp [\frac{%
3\eta Vd}{10\sqrt{\Phi _{0}}}]\left( V^{2}+\frac{\eta V^{3}d}{10\sqrt{\Phi
_{0}}}\right) \right\}  \nonumber
\end{eqnarray}
and

\begin{eqnarray}
G_{(AP)} &=&\frac{2\pi e^{2}}{\hbar }\exp [-1.024d\sqrt{\Phi _{0}}%
][N_{m}^{F}]^{2}\left\{ 2r\exp [\frac{\eta Vd}{2\sqrt{\Phi _{0}}}]+\frac{%
\beta (r+\lambda )}{3}\times \right.  \nonumber \\
&&\times \left[ \frac{\eta dV^{2}}{4\sqrt{\Phi _{0}}}\exp [\frac{\eta Vd}{2%
\sqrt{\Phi _{0}}}]+V\left( \exp [\frac{\eta Vd}{2\sqrt{\Phi _{0}}}]-1\right)
\right] -  \label{gap} \\
&&\qquad \qquad \left. -\beta ^{2}\lambda \exp [\frac{3\eta Vd}{10\sqrt{\Phi
_{0}}}]\left( V^{2}+\frac{\eta V^{3}d}{10\sqrt{\Phi _{0}}}\right) \right\} \
.  \nonumber
\end{eqnarray}

The expressions above can be easily inverted to obtain the resistance, with
the MR defined as 
\begin{equation}
\frac{\Delta R}{R}=\frac{R_{AP}-R_{P}}{R_{AP}}\ .  \label{mr}
\end{equation}
Note that the above definition is limited to $100\%$, since $R_{AP}>R_{P}$.
In the limit $V\rightarrow 0$ we have approximately 
\[
G_{(P)}=\frac{2\pi e^{2}}{\hbar }\exp [-1.024d\sqrt{\Phi _{0}}%
][N_{m}^{F}]^{2}(1+r^{2})\exp [\frac{\eta Vd}{2\sqrt{\Phi _{0}}}] 
\]
and

\[
G_{(AP)}=\frac{2\pi e^{2}}{\hbar }\exp [-1.024d\sqrt{\Phi _{0}}%
][N_{m}^{F}]^{2}(2r)\exp [\frac{\eta Vd}{2\sqrt{\Phi _{0}}}]\ . 
\]
~ With the experimental value of $\Delta R/R$ at zero bias, we can easily
obtain the ratio of the densities of states $r$ at the Fermi level by:

\begin{equation}
r=\frac{1}{1-\left. \frac{\Delta R}{R}\right| _{V=0}}+\ \sqrt{\frac{1}{%
\left( 1-\left. \frac{\Delta R}{R}\right| _{V=0}\right) ^{2}}-1}\ ,
\label{r}
\end{equation}
which does not depend on the barrier parameters. In turn, the barrier height 
$\Phi _{0}$ and thickness $d$ determine the absolute value of the
resistance. Typical values used in our examples are $d=1.0\ nm$ and $\Phi
_{0}=3.0$ $eV$. Estimation of the resistance of such a junction yields
resistance-area products of the order of $RS\approx 3.3\times 10^{4}\ \left[
\Omega \ \mu m^{2}\right] $, where $S$ is the junction area given in $\mu
m^{2}$. This value follows closely the resistance-area scaling obtained for
different junctions in Ref. \onlinecite{gallagher}, with values of the MR
ranging from $16\%$ to $22\ \%$. Representative experimental data of the
tunneling resistance dependence on bias are given in Ref. %
\onlinecite{[2],[3],[7],moodera}. We compare our theoretical calculation
with results presented in Ref. \onlinecite{[7]} at $4.2^{\circ }K$. There,
the {\em zero-bias }MR is approximately of the order of $25\%$, which yields
for the $r$ parameter of (\ref{parameters}) and (\ref{r}) the value $r=2.21$%
. In Fig. 1 we show our theoretical results for the resistance calculated
with formulae (\ref{gp}) and (\ref{gap}). The band structure parameters were
taken with the values $\lambda =0.07$ and $\beta =2.7$, and the tunneling
parameter as $\eta =0.1$. The small value of $\lambda $ depicts a situation
where the majority spin band is saturated at the Fermi level, while the
minority one has a large variation\cite{[8]}. However, the slope of the
resistance near zero bias only depends on the ratio of the densities of
states, in the form 
\begin{equation}
\begin{array}{c}
R_{AP}\approx R_{0}\left( 
{\displaystyle {1 \over 2r}}%
-%
{\displaystyle {1 \over 2r}}%
x\right) \ , \\ 
\\ 
R_{P}\approx R_{0}\left( 
{\displaystyle {1 \over 1+r^{2}}}%
-%
{\displaystyle {1 \over 1+r^{2}}}%
x\right) \ ,
\end{array}
\   \label{resistances}
\end{equation}
where $x=%
{\displaystyle {\eta d \over 2\sqrt{\Phi _{0}}}}%
\left| V\right| $ and $R_{0}=%
{\displaystyle {\exp [1.024d\sqrt{\Phi _{0}}] \over \left( \frac{2\pi e^{2}}{\hbar }\right) [N_{m}^{F}]^{2}}}%
$ is a scale factor related to the absolute resistance. Note that we get a 
{\em zero bias anomaly}, but a good fit with the experiment is only obtained
for the parallel configuration, as in Ref. \onlinecite{[8]}. One can
adequate the theoretical model to a better fit with the data, using more
terms in the Taylor expansion of $W_{(C)}$, or leaving the densities of
states as free parameters\cite{[2]}. However, we interpret the failure of
fitting the data for the AP configuration as a hint that points to the
contribution of an extra mechanism, which affects differently the P and AP
resistances. The linear terms in (\ref{resistances}) cancel out when one
gets the MR, as shown in Fig. 2, along with the experimental data. We pursue
our argument further in the next section, with the inclusion of magnon
inelastic scattering processes in the calculation of the MR.

\section{Magnon-assisted inelastic tunneling}

In this section we consider not only the elastic (spin conserving) processes
but inelastic magnon-assisted contributions to the tunneling current. The
latter are responsible for opening the spin-flip channels, substantially
reducing the MR near zero bias. Magnons are spin-wave excitations \cite{[15]}
which interact with electrons, being emitted or absorbed, thus producing
changes in their energy and allowing for spin-flip scattering. Electrons
accelerated by the electric field relax their energy, producing those
collective excitations at the magnetic electrode interfaces. At low
temperature, only magnon emission processes give a significant contribution
to the resistance. However we analyze in the following the general case,
describing each one of the eight processes associated with emission and
absorption or magnons. There is one extra term related to the overlap
between wave functions of the electrodes, not involving changes in the
number of magnons. This term is proportional to the exchange transmission
coefficient $T^{J}=|t_{kk^{\prime }q}^{J}|^{2}$, resulting in a very similar
formula to the one found for the direct tunneling in the previous section:

\[
I_{(C)}^{N}=\frac{2\pi e}{\hbar }\int d\varepsilon
~(S_{R}^{2}+S_{L}^{2})~T^{J}(\epsilon ,V,d,\Phi _{0})~W_{(C)}(\varepsilon
,V)[f(\varepsilon -V)-f(\varepsilon )]\ . 
\]

Let us consider now the electron tunneling from the left to the right
electrode with the emission of one magnon at the right side interface: 
\begin{eqnarray*}
I_{(C)}^{E1} &=&\frac{2\pi e}{\hbar }\int d\omega \int d\varepsilon
~2S_{R}~T^{J}(\varepsilon ,V,d,\Phi _{0})~\times \\
&&\times ~N_{\downarrow }^{L}(\varepsilon -V+\hbar \omega )N_{\uparrow
}^{R}(\varepsilon )~\rho _{R}^{mag}(\omega )[1+f_{BE}(\omega )]f(\varepsilon
-V+\hbar \omega )[1-f(\varepsilon )]\ ,
\end{eqnarray*}
being $\rho ^{mag}(\omega )$ the density of magnons at the right side
interface and $f_{BE}$ the Bose-Einstein distribution:

\[
f _{BE}=\frac{1}{\exp[\frac{\hbar \omega}{k_B T}]-1} 
\]

An identical expression appears when considering the magnon emission at the
left side interface yielding:

\begin{eqnarray}
I_{(C)}^{E1} &=&\frac{2\pi e}{\hbar }\int d\omega \int d\varepsilon
~2(S_{R}\rho _{R}^{mag}(\omega )+S_{L}\rho _{L}^{mag}(\omega
))~T^{J}(\varepsilon ,V,d,\Phi _{0})~\times  \label{ie1} \\
&&\times ~N_{\downarrow }^{L}(\varepsilon -V+\hbar \omega )N_{\uparrow
}^{R}(E)~[1+f_{BE}(\omega )]f(\varepsilon -V+\hbar \omega )[1-f(\varepsilon
)]\ .  \nonumber
\end{eqnarray}

When the tunneling occurs from right to left with one magnon emission, we
have: 
\begin{eqnarray}
I_{(C)}^{E2} &=&\frac{2\pi e}{\hbar }\int d\omega \int d\varepsilon
~2(S_{R}\rho _{R}^{mag}(\omega )+S_{L}\rho _{L}^{mag}(\omega
))~T^{J}(\varepsilon ,V,d,\Phi _{0})~\times  \label{ie2} \\
&&\times ~N_{\uparrow }^{L}(\varepsilon -V+\hbar \omega )N_{\downarrow
}^{R}(\varepsilon )~[1+f_{BE}(\omega )]f(\varepsilon )[1-f(\varepsilon
-V+\hbar \omega )]\ .  \nonumber
\end{eqnarray}

In turn, for magnon absorption we get:

\begin{eqnarray}
I_{(C)}^{A1} &=&\frac{2\pi e}{\hbar }\int d\omega \int d\varepsilon
~2(S_{R}\rho _{R}^{mag}(\omega )+S_{L}\rho _{L}^{mag}(\omega
))~T^{J}(\varepsilon ,V,d,\Phi _{0})~\times  \label{ia1} \\
&&\times ~N_{\uparrow }^{L}(\varepsilon -V-\hbar \omega )N_{\downarrow
}^{R}(\varepsilon )~[f_{BE}(\omega )]f(\varepsilon -V-\hbar \omega
)[1-f(\varepsilon )]  \nonumber
\end{eqnarray}
and

\begin{eqnarray}
I_{(C)}^{A2} &=&\frac{2\pi e}{\hbar }\int d\omega \int d\varepsilon
~2(S_{R}\rho _{R}^{mag}(\omega )+S_{L}\rho _{L}^{mag}(\omega
))~T^{J}(\varepsilon ,V,d,\Phi _{0})~\times  \label{ia2} \\
&&\times ~N_{\downarrow }^{L}(\varepsilon -V-\hbar \omega )N_{\uparrow
}^{R}(\varepsilon )~[f_{BE}(\omega )]f(\varepsilon )[1-f(\varepsilon
-V-\hbar \omega )]\ .  \nonumber
\end{eqnarray}

The total current due to one magnon exchange is then:

\begin{equation}
I_{mag}=I_{(C)}^{N}+I_{(C)}^{E1}-I_{(C)}^{E2}+I_{(C)}^{A1}-I_{(C)}^{A2}
\label{imag}
\end{equation}

Typical IET magnon spectra are shown by Ando and coworkers in Ref. %
\onlinecite{[1]}. They display a strong peak around $12-20$ $mV$ and a rapid
decrease for energies below the peak, due probably to a low energy cutoff,
with a vanishing magnon density of states at very small energies.
Introducing this low energy cutoff in the magnon spectrum, and taking the
low temperature limit $T\rightarrow 0^{\circ }K$, we get $f_{BE}\rightarrow
0 $ for the Bose-Einstein distribution. This limit excludes the absorption
terms in (\ref{imag}), leaving only the emission contributions to the total
current: 
\begin{eqnarray}
I_{mag} &=&\frac{4\pi e}{\hbar }\int d\omega \int_{0}^{V-\hbar \omega
}d\varepsilon \left\{ T^{J}(\varepsilon ,V,d,\Phi _{0})(S_{R}\rho
_{R}^{mag}(\omega )+S_{L}\rho _{L}^{mag}(\omega ))~\times \right.  \nonumber
\\
&&\qquad \qquad \qquad \times N_{\downarrow }^{L}(\varepsilon -V-\hbar
\omega )N_{\uparrow }^{R}(\varepsilon )\ \Theta \left( V-\hbar \omega
\right) -  \nonumber \\
&&-T^{J}(\varepsilon ,V,d,\Phi _{0})(S_{R}\rho _{R}^{mag}(\omega )+S_{L}\rho
_{L}^{mag}(\omega ))~\times  \label{imag2} \\
&&\qquad \qquad \qquad \left. \times N_{\downarrow }^{R}(\varepsilon
)N_{\uparrow }^{L}(\varepsilon -V-\hbar \omega )\ \Theta \left( \hbar \omega
-V\right) \right\} +I_{(C)}^{N}\ ,  \nonumber
\end{eqnarray}
being $\Theta (x)$ the step function.

One can use as the magnon dispersion relation a simple isotropic parabolic
dependence,{\em \ i.e.}, $\hbar \omega =E_{m}(q/q_{m})^{2}$, where $E_{m}$
is related to the Curie temperature by the mean field approximation $%
E_{m}=3k_{B}T_{C}/(S+1)$, and $q_{m}$ is the radius of the first Brillouin
zone\cite{[7]}. In other words $E_{m}$ is the maximum magnon energy (high
energy cutoff)\cite{[1]}. Considering the above discussion, assuming
identical ferromagnetic electrodes, and after some mathematical
simplifications, one finally gets the conductance in the form: 
\[
G_{C}=G_{C}^{d}+G_{C}^{mag} 
\]
where $G_{C}^{d}$ is given by (\ref{gp}) and (\ref{gap}), for P and AP
alignment respectively, and $G_{C}^{mag}$ is shown below:

\begin{equation}
G_{P}^{mag}=\frac{2\pi e^{2}}{\hbar }T^{J}(V)\left[ 2S^{2}W_{(P)}+\Lambda
(V)SW_{(AP)}\right] \ ,  \label{gpmag}
\end{equation}
and

\begin{equation}
G_{AP}^{mag}=\frac{2\pi e^{2}}{\hbar }T^{J}(V)\left[ 2S^{2}W_{(AP)}+2\Lambda
(V)SW_{(P)}\right] \ ,  \label{gapmag}
\end{equation}
with $S=\left( S^{R}+S^{L}\right) /2$ and 
\[
\Lambda \left( V\right) =\left\{ 
\begin{array}{l}
V/E_{m}\quad {\rm for\ }V<E_{m}\ , \\ 
2-E_{m}/V\quad {\rm for\ }V>E_{m}\ .
\end{array}
\right. 
\]
The functions $W_{(P)}$ and $W_{(AP)}$ in (\ref{gpmag}) and (\ref{gapmag})
have been evaluated near zero bias, using formulae (\ref{parallel}) and (\ref
{antiparallel}) respectively, substituting $\varepsilon $ by the constant
value $0.1\ eV$. We point out that magnon processes significantly contribute
to the conductance only for voltages below $100\ mV$, and so we can consider 
$W_{(P)}$ and $W_{(AP)}$ as almost constant under the integration sign. The
exchange tunneling coefficient $T^{J}$ generally is one or two orders of
magnitude smaller than the direct coefficient. We found excellent agreement
between our theory and the experimental data using the same set of
parameters of Fig. 1 for the tunneling barrier and the electronic structure,
spin $S=3/2$ and $T^{d}/T^{J}=37$ for the ratio of the direct tunneling to
the exchange tunneling coefficient. The magnon cutoff $E_{m}$ was taken to
be $\ 90\ meV$. The results are shown in Fig. 3 for the resistances and in
Fig. 4 for the corresponding MR. Clearly, the AP configuration is more
sensible to the magnon contribution, since the current for that
configuration is weighted by the product $N_{M}^{L}N_{M}^{R}$, which is much
bigger than the factors $N_{m}^{L}N_{M}^{R}$ or $N_{M}^{L}N_{m}^{R}$ which
appear in the P current, with the indices $m$ and $M$ referring to {\em %
minority} and {\em majority }spin bands (compare Fig. 1 and Fig. 3).
Obviously, minor differences between theory and experimental data come from
the fact that we are using a very simplified model for the band structure
and the magnon dispersion relation. We comment on these results in the next
section.

\section{Conclusions}

We have presented a consistent study of the voltage dependence of the \
`giant' magnetoresistance in ferromagnetic tunneling junctions. Our approach
includes: {\em a) }lowering of the effective barrier height with the applied
voltage; {\em b) }different variations of the density of states for each
spin band with voltage; and {\em c)} magnon assisted inelastic tunneling
near zero bias. We found that taking into account all those effects is
essential to fully explain experimental results at low temperature for the
voltage range between $0$ and $500$ $mV$. We have also clarify the role of
the different parameters used in the theory: some of them ($d,\Phi _{0},\eta 
$) determine the absolute value of the resistance at zero bias, which in
turn is a scale factor in the theory; a different set, related to the band
structure ($r,\beta ,\lambda $), mainly monitors the global behavior with
voltage and the value of the junction MR. To adjust our results with
selected experimental data, we have taken $\beta ,\lambda >0$, but as shown
in Ref. \onlinecite{[8]}, this scenario is not unique and depends on the
topology of the bands that contribute to the current; and finally, the
behavior near zero bias ({\em zero bias anomaly}), with a rapid decrease of
the resistance for the AP configuration up to $100\ mV$, is ascribed to
magnon-assisted tunneling. Our estimation of $T^{d}/T^{J}\sim 40$ seems to
be more realistic than previous estimations\cite{[7]}. We note that the
latter is the only adjustable parameter to fit the voltage dependence below $%
100\ mV$ (for both, P and AP configurations). Our calculation is in
excellent agreement with the experimental data (see Fig. 3 and 4).

Temperature effects are not discussed in this paper. As shown in Section IV,
only magnon emission processes are included at low temperature ($%
T\rightarrow 0$). At finite temperature, we expect a decrease of the
resistance near zero bias, due to one-magnon-absorption assisted tunneling 
\cite{dartora}. The above should be superimposed to the thermal smearing in
the Fermi-Dirac distribution of tunneling electrons\cite{[3]}.

\acknowledgments
The authors would like to acknowledge partial financial support from {\em %
Funda\c{c}\~{a}o do Amparo \`{a} Pesquisa do Estado de S\~{a}o Paulo}
(FAPESP, SP, Brazil), through the project \# 2002/09895-6.


\figure{Fig.1  } Resistance as a function of the voltage bias for the AP and
P configurations: the experimental results (dotted line and symbols) are
taken from Ref. \onlinecite{[7]} and the theoretical ones (solid lines) are
calculated with formulae (\ref{gp}) and (\ref{gap}), using the following
parameters: $d=1.0\ nm$, $\Phi _{0}=3.0\ eV$, $N_{m}^{F}=1.0$ in normalized
units, $r=2.21$, $\lambda =0.07$, $\beta =2.7$, and $\eta =0.1$. The
resistances are given in arbitrary units, normalized to the peak value at
zero bias.

\figure{Fig.2  } Magnetoresistance as a function of voltage. Parameters are
kept the same as in Fig.1.

\figure{Fig.3  } Resistance, in arbitrary units, as a function of the
voltage bias for the AP and P configurations: the experimental results
(dotted line and symbols) are taken from Ref. \onlinecite{[7]} and the
theoretical ones (solid lines) include magnon-assisted tunneling. Parameters
are kept the same as in Fig. 1, with the addition of $T^{d}/T^{J}=37$.

\figure{Fig.4  } Magnetoresistance as a function of voltage. Parameters are
kept the same as in Fig.3.


\begin{references}
\bibitem{[1]}  Y. Ando, J. Murai, H. Kubota and T. Miyazaki, J. Appl. Phys. 
{\bf 87}, 5209 (2000).

\bibitem{[2]}  X. H. Xiang, T. Zhu, J. Du, G. Landry and J. Q. Xiao, Phys.
Rev. B {\bf 66}, 174407 (2002).

\bibitem{[3]}  J. J. Akerman, I. V. Roushchin, J. M. Slaughter, R. W. Dave
and I. K. Schuller, Europhys. Lett. {\bf 63}, 104 (2003).

\bibitem{[4]}  F. Montaigne, J. Nassar, A. Vaur\`{e}s, F. Nguyen Van Dau, F.
Petroff, A. Schuhl and A. Fert, Appl. Phys. Lett. {\bf 73}, 2829 (1998).

\bibitem{[5]}  T. Miyazaki and N. Tezuka, J. Magn. Magn. Mater. {\bf 139},
L231 (1995).

\bibitem{[18]}  G. G. Cabrera and L. M. Falicov, Phys. Status Solidi B {\bf %
61}, 539 (1974); Phys. Rev. B {\bf 11}, 2651 (1975).

\bibitem{[6]}  M. Julli\`{e}re, Phys. Lett. A {\bf 54}, 225 (1975).

\bibitem{[7]}  S. Zhang, P. M. Levy, A. C. Marley and S. S. P. Parkin, Phys.
Rev. Lett. {\bf 79}, 3744 (1997).

\bibitem{moodera}  J. S. Moodera, J. Nowak, R. J. M. van de Veerdonk, Phys.
Rev. Lett. {\bf 80}, 2941 (1998).

\bibitem{[8]}  G. G. Cabrera and N. Garc\'{\i }a, Appl. Phys. Lett. {\bf 80}%
, 1782 (2002).

\bibitem{[10]}  J. G. Simmons, J. Appl. Phys. {\bf 34}, 1793 (1963).

\bibitem{other}  M. Sharma, S. W. Wang, and J. H. Nickel, Phys. Rev. Lett. 
{\bf 82}, 616 (1999).

\bibitem{[11]}  P. LeClair, J. T. Kohlhepp, H. J. M. Swagten and W. J. M. de
Jonge, Phys. Rev. Lett. {\bf 86}, 1066 (2001).

\bibitem{nico}  N. Garc\'{\i }a, Appl. Phys. Lett. {\bf 77}, 1351 (2000).

\bibitem{[12]}  D. K. Ferry and S. M. Goodnick, {\em Transport in
Nanostructures} (Cambridge University Press, Cambridge, 1997); Y. Imry, {\em %
Introduction to Mesoscopic Physics} (Oxford University Press, Oxford, 1997).

\bibitem{[15]}  D. C. Mattis, {\em The Theory of Magnetism} (Harper and Row
Publishers, New York, 1965); C. Kittel, {\em Quantum Theory of Solids} (John
Wiley, New York, 1963).

\bibitem{gallagher}  W. J. Gallagher, S. S. P. Parkin, Yu Lu, X. P. Bian, A.
Marley, K. P. Roche, R. A. Altman, S. A. Rishton, C. Jahnes, T. M. Shaw, and
Gang Xiao, J. Appl. Phys. {\bf 81}, 3741 (1997).

\bibitem{dartora}  C. A. Dartora and G. G. Cabrera, unpublished.
\end{references}
\end{document}